\documentclass[journal]{IEEEtran}
\usepackage{cite}
\usepackage{amsmath,amssymb,amsfonts,amsthm}
\usepackage{BOONDOX-calo}
\usepackage[linesnumbered, vlined]{algorithm2e}
\usepackage{graphicx}
\usepackage{dblfloatfix}
\usepackage{textcomp}
\usepackage{xcolor}
\usepackage[acronym]{glossaries}
\usepackage{float}
\usepackage{subcaption}
\floatstyle{ruled}
\newfloat{model}{thp}{lop}
\floatname{model}{Model}
\def\BibTeX{{\rm B\kern-.05em{\sc i\kern-.025em b}\kern-.08em
    T\kern-.1667em\lower.7ex\hbox{E}\kern-.125emX}}
\makeatletter
\newcommand{\removelatexerror}{\let\@latex@error\@gobble}
\makeatother

\newtheorem{definition}{Definition}
\newtheorem{theorem}{Theorem}
\newtheorem{proposition}{Proposition}

\newtheorem{corollary}{Corollary}

\newcommand{\B}{\mathcal{B}}
\newcommand{\C}{\mathcal{C}}
\newcommand{\CF}{\mathcal{C}^\mathcal{B}}
\newcommand{\CN}{\mathcal{C}^\mathcal{N}}

\newcommand{\E}{\mathcal{E}}
\newcommand{\EN}{\mathcal{E}^\mathcal{N}}
\newcommand{\Es}{\mathcal{E^{S}}}

\newcommand{\f}{\mathcal{f}}
\newcommand{\G}{\mathcal{G}}
\renewcommand{\H}{\mathcal{H}}

\newcommand{\N}{\mathcal{N}}

\newcommand{\T}{\mathcal{T}}

\newcommand{\V}{\mathcal{V}}
\newcommand{\Vr}{\mathcal{V}^{\text{ref}}}
\newcommand{\Lo}{\mathcal{L}}
\newcommand{\mgtd}[1]{\lvert #1 \rvert}

\newcommand{\mcdsr}{Model \ref{mod:cdsr}-\gls{cdsr}\xspace}

\newcommand{\mvr}{\gls{mv}-Rural\xspace}
\newcommand{\mvc}{\gls{mv}-Comm\xspace}
\newcommand{\mvs}{\gls{mv}-Semiurb\xspace}

\newcommand{\ts}{\textstyle\sum}
\newcommand{\set}[1]{\{#1\}}
\newcommand{\vts}[1]{\lvert #1\rvert}

\begin{document}

\newacronym{bnb}{B\&B}{branch-and-bound}%
\newacronym{mv}{MV}{medium voltage}%
\newacronym{hv}{HV}{high voltage}%
\newacronym[\glsshortpluralkey={DAPs}, \glslongpluralkey={day-ahead prices}]{dap}{DAP}{day-ahead prices}%
\newacronym[\glsshortpluralkey={DoFs}, \glslongpluralkey={degrees of freedom}]{dof}{DoF}{degree of freedom}%
\newacronym[\glsshortpluralkey={BZNs}, \glslongpluralkey={bidding zones}]{bzn}{BZN}{bidding zone}%
\newacronym[\glsshortpluralkey={MV-Ss}, \glslongpluralkey={MV-substations}]{mvs}{MV-S}{MV-substation}%
\newacronym[\glsshortpluralkey={HV-Ss}, \glslongpluralkey={HV-substations}]{hvs}{HV-S}{HV-substation}%
\newacronym[\glsshortpluralkey={CBs}, \glslongpluralkey={cycle bases}]{cb}{CB}{cycle basis}%
\newacronym{fsc}{FSC}{fundamental system of cycles}%
\newacronym[\glsshortpluralkey={CEs}, \glslongpluralkey={cycle-edges}]{ce}{CE}{cycle-edge}%
\newacronym[\glsshortpluralkey={UFCEs}, \glslongpluralkey={unique fundamental cycle edges}]{ufce}{UFCE}{unique fundamental cycle edge}%
\newacronym[\glsshortpluralkey={UESs}, \glslongpluralkey={unique edge sets}]{ues}{UES}{unique edge set}%
\newacronym[\glsshortpluralkey={rUESs}, \glslongpluralkey={reduced unique edge sets}]{rues}{rUES}{reduced unique edge set}%
\newacronym{dsr}{DSR}{distribution system reconfiguration}%
\newacronym{rdsr}{R-DSR}{restrictable distribution system reconfiguration}%
\newacronym{cdsr}{C-DSR}{complete distribution system reconfiguration}%
\newacronym{rrdsr}{RR-DSR}{relaxed radiality distribution system reconfiguration}%
\newacronym{hdsr}{H-DSR}{heuristic distribution system reconfiguration}%
\newacronym[\glsshortpluralkey={FCs}, \glslongpluralkey={fundamental cycles}]{fc}{FC}{fundamental cycle}%
\newacronym{mip}{MIP}{mixed-integer program(-ming)}%
\newacronym[\glsshortpluralkey={MILPs}, \glslongpluralkey={mixed-integer linear programs}]{milp}{MILP}{mixed-integer linear program(-ming)}%
\newacronym[\glsshortpluralkey={MIQPs}, \glslongpluralkey={mixed-integer quadratic programs}]{miqp}{MIQP}{mixed-integer quadratic program}%
\newacronym[\glsshortpluralkey={MISOCs}, \glslongpluralkey={mixed-integer second-order cone programs}]{misoc}{MISOC}{mixed-integer second-order cone program}%
\newacronym[\glsshortpluralkey={MINLPs}, \glslongpluralkey={mixed-integer non-linear programs}]{minlp}{MINLP}{mixed-integer non-linear program}%
\newacronym{nlp}{NLP}{non-linear program}%
\newacronym{nlpr}{NLP-R}{NLP relaxation}%
\newacronym{lns}{LNS}{large neighborhood search}%
\newacronym{mcb}{MCB}{minimum cycle basis}%
\newacronym{pf}{PF}{power flow}%
\newacronym{acpf}{ACPF}{alternating current power flow}%
\newacronym{soc}{SOC}{second-order cone}%
\newacronym{dso}{DSO}{distribution system operator}%
\newacronym{dsoe}{DSO-E}{DSO Entity}%
\newacronym{entsoe}{ENTSO-E}{European Network of Transmission System Operators for Electricity}%
\newacronym{ev}{EV}{electric vehicle}%
\newacronym[\glsshortpluralkey={RESs}, \glslongpluralkey={renewable energy sources}]{res}{RES}{renewable energy source}%
\newacronym{it}{IT}{information technology}
\newacronym{snb}{SNB}{Stromnetz Berlin}
\newacronym[\glsshortpluralkey={STs}, \glslongpluralkey={spanning trees}]{st}{ST}{spanning tree}
\newacronym{us}{US}{United States}
\newacronym{ub}{UB}{upper bound}

\title{On Loss-Minimal Radial Topologies in MV Systems}
\author{Anton Hinneck, \IEEEmembership{Member, IEEE}, Mathias Duckheim, 
Michael Metzger, and
Stefan Niessen
\thanks{Anton Hinneck (corresponding author) is with Technische Universität Darmstadt, Electrical Engineering and Information Technology, Landgraf-Georg-Str. 4, Darmstadt, 64283, Germany, and distribution system operator Stromnetz Berlin (e-mail: anton.hinneck@stromnetz-berlin.de).}
\thanks{Mathias Duckheim is with Siemens AG, Foundational Technologies, Schuckertstr. 2, Erlangen, 91058, Germany (e-mail: mathias.duckheim@siemens.com).}
\thanks{Michael Metzger is with Siemens AG, Foundational Technologies, Otto-Hahn-Ring 6, 81739, Munich (e-mail: michael.metzger@siemens.com).}
\thanks{Stefan Niessen is with Technische Universität Darmstadt, Electrical Engineering and Information Technology, Landgraf-Georg-Str. 4, Darmstadt, 64283, Germany
and Siemens AG, Foundational Technologies, Schuckertstr. 2, Erlangen, 91058, Germany (e-mail: stefan.niessen@siemens.com).}
}
\maketitle

\begin{abstract}
\Gls{dsr} means optimizing the topology of a distribution grid using switching actions. Switching actions are a \gls{dof} available to \glspl{dso}, e.g. to manage planned and unplanned outages. 
\Gls{dsr} is a NP-hard combinatorial problem. Finding good or even optimal solutions is computationally expensive. While transmission and high-voltage grids are generally operated in a meshed state, \gls{mv} distribution systems are commonly operated as radial networks even though meshed operation would be supported. This improves resilience because faults can be isolated more easily keeping the rest of the system operational and minimizing impact on customers. We propose an AC \gls{dsr} formulation and benchmark it against a common formulation from the literature. Our results indicate that additional acyclicity constraints can significantly improve solver performance.
\end{abstract}


\section*{Nomenclature}
\addcontentsline{toc}{section}{Nomenclature}

\begin{IEEEdescription}[\IEEEusemathlabelsep\IEEEsetlabelwidth{$~~~~~~~~$}]
\item[$\N$] A graph/power system topology $\set{\E,\V}$
\item[$\H$] A (spanning) subgraph in $\N$
\item[$\T$] A (spanning) tree in $\N$
\item[$\V^{(\text{ref})}$] Set of (reference) buses
\item[$\E^{(\mathcal{S})}$] Set of (switchable) power lines
\item[$\G_{(f)}$] Set of generators (at bus $f$)
\item[$\Lo_f$] Set of loads at bus $f$
\item[$\CN$] Set of all cycles in $\N$
\item[$\EN_i$] Edge set of i-th cycle in $\CN$
\item[$\B$] A \gls{cb} of $\N$
\item[$\CF_k$] The k-th fundamental cycle in $\B$
\item[$\mathcal{f}$] Objective value
\item[$p^{\text{G}}_g/q^{\text{G}}_g$] Real/Reactive power injection of source $g$
\item[$p_{ft}/q_{ft}$] Real/reactive power flow from bus $f$ to $t$
\item[$z_{ft}$] Switching status of line $ft$
\item[$v^m_f$] Voltage magnitude at bus $f$
\item[$\theta_f$] Voltage angle at bus $f$
\item[$\Delta \theta_{tf}$] Voltage angle difference between buses $t$ and $f$ 
\item[$p_l/q_l$] Real/reactive power draw of load $l$
\item[$v^{m,ref}_f$] Measured voltage magnitude at reference bus $f$
\item[$y_{ft}$] Admittance of line $ft$
\item[$g_{ft}/b_{ft}$] Conductance/susceptance of line $ft$
\item[$\underline{v}^m_{f}/\overline{v}^m_{f}$] Minimum/maximum voltage magnitude at bus $f$
\item[$\gamma^{v/s}$] Absolute violations of voltage/power limits
\item[$\overline{s}_{ft}$] Maximum power rating of line $ft$
\end{IEEEdescription}

\begin{model*}[t]
\caption{R-DSR \hfill [MINLP]}
\label{mod:rdsr}
\begin{subequations}
\begin{alignat}{2}
    &\underset{v^m,\theta,p^{\text{G}},q^{\text{G}},p,q,z}{\mathrm{min~}}\f=\ts_{g\in\G}p^{\text{G}}_{g}\label{dsr:obj}\\
    \ts_{g\in\G_{f}}p^{\text{G}}_{g} &= \ts_{t\in\E_f} p_{ft} + \ts_{t\in\E_f} p_{tf} + \ts_{l\in\Lo_{f}}p_{l},\quad&&\forall~ f\in\V\label{dsr:mcr}\\
    \ts_{g\in\G_{f}}q^{\text{G}}_{g}&=\ts_{t\in\E_f} q_{ft}+\ts_{t\in\E_f} q_{tf} + \ts_{l\in\Lo_{f}}q_{l},\quad&&\forall~ f\in\V\label{dsr:mci}\\
    %
    p_{ft} &={\color{white} -}\big(\mgtd{v_f}^2 (g_{ft}+g^{sh}_{ft})
    -v^m_{f}v^m_{t}(g_{ft}\cos(\Delta \theta_{ft}) + b_{ft}\sin(\Delta \theta_{ft}))\big)z_{ft},\quad&&\forall~ ft,tf\in\Es\label{dsr:pfr}\\
    q_{ft} &=-\big(\mgtd{v_f}^2 (b_{ft}+b^{sh}_{ft})
    + v^m_{f}v^m_{t}(b_{ft}\cos(\Delta \theta_{ft})-g_{ft}\sin(\Delta \theta_{ft}))\big)z_{ft},\quad&&\forall~ ft,tf\in\Es\label{dsr:qfr}\\
    %
    %
    p_{ft} &={\color{white} -}\mgtd{v_f}^2 (g_{ft}+g^{sh}_{ft})
    -v^m_{f}v^m_{t}(g_{ft}\cos(\Delta \theta_{ft}) + b_{ft}\sin(\Delta \theta_{ft})),\quad&&\forall~ ft,tf\in\E\backslash\Es\label{dsr:pfrF}\\
    q_{ft} &=-\mgtd{v_f}^2 (b_{ft}+b^{sh}_{ft})
    + v^m_{f}v^m_{t}(b_{ft}\cos(\Delta \theta_{ft})-g_{ft}\sin(\Delta \theta_{ft})),\quad&&\forall~ ft,tf\in\E\backslash\Es\label{dsr:qfrF}\\
    %
    %
    v^m_{f} &= v^{m,ref}_{f} &&\forall~ f\in\Vr\label{dsr:ref}
    %
    %
\end{alignat}
\end{subequations}
\end{model*}
\section{Introduction}
\label{sec:introduction}
\begin{figure}[ht]
\begin{center}
    \begin{subfigure}[b]{0.24\linewidth}
        \centering
        \includegraphics[width=0.7\linewidth]{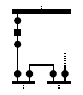}
        \caption{Radial feed}
        \label{fig:topRad}
    \end{subfigure}
    \begin{subfigure}[b]{0.24\linewidth}
        \centering
        \includegraphics[width=0.7\linewidth]{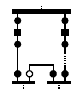}
        \caption{Open ring}
        \label{fig:topOpenRing}
    \end{subfigure}
    \begin{subfigure}[b]{0.24\linewidth}
        \centering
        \includegraphics[width=0.7\linewidth]{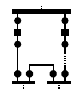}
        \caption{Closed ring}
        \label{fig:topRing}
    \end{subfigure}
    \caption{This figure conceptually displays distribution system topologies, based on a German national standard \cite{4110}.}
    \label{fig:topos}
\end{center}
\end{figure}

The topology of an electrical network can be represented by a graph $\N=\{\V,\E\}$ where $\V$ denotes a set of vertices and $\E$ a set of edges. Improving the electrical state of the network requires power flow expressions as well as integer variables to model switches in optimization models. Power flow equations are non-convex, which makes them difficult to solve. The same applies to integrality constraints on variables. While specialized \gls{minlp} solvers exist, %
the resulting problems may not be tractable computationally for real-world systems. While several approaches have been proposed using relaxed \gls{pf} approximations that yield \glspl{milp}, \glspl{misoc} or \glspl{miqp}, few consider the exact \gls{minlp} \cite{Raju2008-vy, De_Bonis2014-nq, Lavorato2012-op, Capitanescu2015-qe}. \glspl{minlp} provide no general guarantees for global optimality but are the most accurate in modeling many real-world systems. For such problems Juniper \cite{juniper}, the solver used on problems in this work, is in fact a heuristic. While some convex \gls{pf} approximations like the \gls{soc} approximation are only exact for radial system topologies, which also includes open rings, and linear approximations generally provide inaccurate results in all dimensions \cite{Taylor2018}, the exact \gls{pf} model used in this work provides accurate results independently of system topologies. Common distribution system topologies are displayed in Figure \ref{fig:topos}.
The biggest drawback of exact \gls{pf} constraints is their computational burden. Besides \gls{pf}, radiality of solutions is another challenge.
The radiality condition used in this work has only been used in a few other publications \cite{Borghetti2012-yc, Arif2020-te, Pang2023-ef}, none of which considers the exact AC load flow constraints.
\newline
The contributions in this work are three-fold. We propose
\begin{itemize}
    \item A model for \gls{dsr} considering exact AC load flow constraints and the radiality condition stated in Theorem \ref{thm:tree}, 2), using the set of all cycles,
    \item A theoretical result to eliminate integer variables and reduce complexity, as well as
    \item A computational comparison against a common radiality condition used in the literature \cite{Raju2008-vy, De_Bonis2014-nq, Lavorato2012-op, Capitanescu2015-qe}.
\end{itemize}

\section{Models for loss-minimal radial topologies}
\label{sec:prerequisites}
The \gls{dsr} model presented is called \gls{cdsr}. \Gls{rrdsr} is used as a benchmark. Both models are derived from \gls{rdsr}, which is similar to the model proposed in \cite{Hinneck2023}. In \gls{rdsr}, however, an AC \gls{pf} formulation based on \cite{Coffrin2018} is used, which is applicable to single-phase or symmetric three-phase systems.
In all models, consumers in $\Lo$ are modeled using the passive sign convention and power sources in $\G$, which are connections to an external grid in this work, using the active sign convention. This means that power injected has a positive sign at buses with power sources and a negative sign at buses where only consumers or prosumers are placed. For power draw, the signs are respectively reversed.
A measured voltage value is set at the so-called reference bus by constraint \eqref{dsr:ref}. This voltage is determined in practice by a measurement device in the substation. Power injected into branches is modeled using equalities \eqref{dsr:pfrF}-\eqref{dsr:qfrF}. To improve readability we define $\Delta \theta_{ft}:\theta_f-\theta_t$. While these equalities apply for all branches $ft\in\E$ in a generic \gls{pf} formulation, here they only apply for those lines that are not subject to switching, i.e. $ft\in\E\backslash\Es$. 
For every line $ft\in\Es$ power flow constraints only apply if the modeled switch is closed, i.e. $z_{ft}=1$. Else, one has $p_{ft}=q_{ft}=p_{tf}=q_{tf}=z_{ft}=0$.
The power injection is minimized by objective function \eqref{dsr:obj}. It is well-known that minimizing injection minimizes power losses \cite{Taylor2018}.
%
%
\subsubsection{Safety ratings}
To not damage equipment or put personnel at risk voltage \eqref{dsr:vm} and power limits \eqref{dsr:pl1} must be respected.
\begin{alignat}{2}
\underline{v}^m_{f} &\leq v^m_{f} \leq \overline{v}^m_{f},\quad &&\forall~ f\in\V\label{dsr:vm}\\
p_{ft}^2 &+ q_{ft}^2 \leq \overline{s}_{ft}^2,\quad &&\forall~ ft\in\E\label{dsr:pl1}
\end{alignat}
The models in this work are relaxed by removing \eqref{dsr:vm} and \eqref{dsr:pl1} from the \glspl{minlp}' constraint sets, evaluating their violation after optimizations. This has practical merit as minimal violations of constraints \eqref{dsr:vm} \& \eqref{dsr:pl1} are admissible in system operation for short durations. To make documentation concise, terms $\gamma^v$ and $\gamma^s$ are defined in \eqref{dsr:gv} and \eqref{dsr:gs}.
These terms are 0 if the constraints are not violated and are otherwise equal to the constraint violation.
\begin{alignat}{2}
\gamma^v_f&:=\begin{cases}0,~\text{if}~\eqref{dsr:vm} \\ \text{max}(\underline{v}^m_{f}-v^m_{f}, v^m_{f}-\overline{v}^m_{f}),~\text{if}~\neg\eqref{dsr:vm}\end{cases}\label{dsr:gv}\\
\gamma^s_{ft}&:=\begin{cases}0,~\text{if}~\eqref{dsr:pl1} \\ (p_{ft}^2 + q_{ft}^2)-\overline{s}_{ft}^2,~\text{if}~\neg\eqref{dsr:pl1}\end{cases}\label{dsr:gs}
\end{alignat}

\subsection{\Acrlong{cdsr}}
\label{sec:cdsr}
\Gls{cdsr} implicitly considers all \gls{st} and
requires $\CN$ as an input.
To enumerate all cycles in an undirected, unweighted graph Gibbs' algorithm described in \cite{Gibbs69} was implemented.  In this model radiality is enforced by the acyclicity-based \gls{st} definition in 2), Theorem \ref{thm:tree}. Radiality, switching actions and non-convex, non-linear AC load flow equations are jointly considered.
\begin{model}
    \caption{\Gls{cdsr}($\set{\E(\CN_1),\dots,\E(\CN_k)}$) \hfill [\Gls{minlp}]}
    \label{mod:cdsr}
    \begin{subequations}
    \begin{alignat}{2}
        \eqref{dsr:obj}&-\eqref{dsr:ref},~~&&\text{with}~\Es=\E\label{cdsr:complete}\\
        \ts_{(f,t)\in\EN_{k}} z_{ft} &\leq \vts{\E(\CN_k)}-1,~~&&\forall~k\in\{1,\dots,\vts{\CN}\}\label{cdsr:acyclic}\\
        \ts_{(f,t)\in\E} z_{ft} &= \vts{\V} - 1\label{cdsr:nm1}
    \end{alignat}
    \end{subequations}
\end{model}
By \eqref{cdsr:complete} all power lines are subject to switching in the resulting problem. Moreover, constraints \eqref{cdsr:acyclic} guarantee that the resulting operational topology is acyclic because every cycle is at least missing one edge, while \eqref{cdsr:nm1} ensures that any of the resulting topologies have $\vts{\V_{\N}}-1$ edges. By Theorem \ref{thm:tree}, the resulting graph is a tree. By definitions \ref{def:subg} and \ref{def:spt} the result is a \gls{st}. $\V$ stays unchanged, as only lines are subject to switching.
The efficiency of \mcdsr can be improved by reducing the number of integer variables. $\Es=\E$ may not be required.
\begin{corollary}
\label{cor:switchable}
Line $e$ can only be de-energized in a \mcdsr or \gls{rrdsr} instance, iff $e\in\textstyle\bigcup_{\forall k} \EN_k$.
\end{corollary}
Applying Corollary \ref{cor:switchable}, constraints \eqref{cdsr:complete} can be constructed with $\Es=\{e\in\E|e\in\textstyle\bigcup_{\forall k} \EN_k\}$.
Then, constraint \eqref{cdsr:nm1} must be reformulated to match 
\begin{equation}
\vts{\E} - \vts{\Es} + \ts_{(f,t)\in\Es} z_{ft} = \vts{\V} - 1.
\end{equation}
This is because the number of lines now depends on the optimization result. From all lines $\vts{\E}$, the number of switchable lines $\vts{\Es}$ is deducted to then add the number of lines energized by the optimizer $\ts_{(f,t)\in\Es} z_{ft}$.
\newline
Following Corollary \ref{cor:ces}, $\textstyle\bigcup_{\forall k} \EN_k$ in Corollary \ref{cor:switchable} can be obtained given a \gls{cb}, which is less expensive to obtain than the set of all cycles by full enumeration.
\begin{corollary}
\label{cor:ces}
It holds that $\textstyle\bigcup_{i=1}^{\lvert\C^\N\rvert}\E(\C^\N_i)=\textstyle\bigcup_{k=1}^{\beta(\N)}\E(\C^\B_k)$.
\end{corollary}
These results are illustrated in Figure \ref{fig:cycleEdges}. Edges highlighted in black are in no cycle of the graph, which can be verified based on all cycles displayed in Figure \ref{fig:ruralGibbs}. They are, hence, no cycle-edges as defined by Definition \ref{def:cycleEdge}.
By Proposition \ref{prop:conn}, removing any of these edges necessarily disconnects the graph. This can be visually confirmed. Thus, one can permanently exclude or close all switches associated with non-cycle edges, as stated in Corollary \ref{cor:switchable}.
\setcounter{figure}{0}
\begin{figure}[ht]
\centering
    \includegraphics[width=0.4\linewidth]{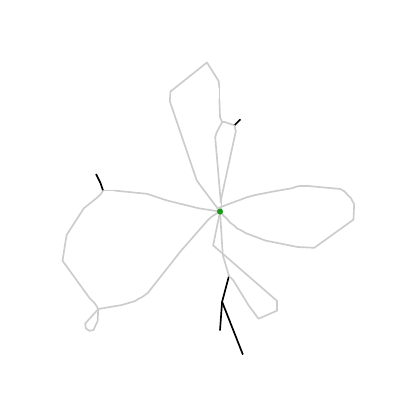}
    \label{fig:ces}
\caption{This plot shows all cycle-edges in gray, identified using Corollary \ref{cor:ces}.}
\label{fig:cycleEdges}
\end{figure}
An efficient way to identify all cycle edges is stated in Corollary \ref{cor:ces} using the graph's \gls{cb}. Instead of enumerating all cycles to form the union of all their edges, it suffices to generate a \gls{cb} and then form the union of edges of its elements. The cycles in subfigures \ref{fig:r_ac1}-\ref{fig:r_ac6} conveniently pose a \gls{cb} of MV-Rural's graph. Assume that cycle \ref{fig:r_ac1} with all its edges was not in that \gls{cb}. Then, this cycle could not be produced as the symmetric difference of any of the remaining basis elements, whereby they no longer form a \gls{cb}, hence the Corollary.
\subsection{\Acrlong{rrdsr}}
\label{sec:refmod}
The reference model used, called \gls{rrdsr}, is equivalent to \gls{cdsr} omitting constraints \eqref{cdsr:acyclic}. This model produces \glspl{st} based on Theorem \ref{thm:tree}, 3), if the system data enforces connectedness. Limitations are discussed in \cite{Lavorato2012-op}. \gls{cdsr} is only in so far dependent on system data that a \gls{st} must exist, which makes it more robust.

\setcounter{figure}{3}
\begin{figure*}[b]
    \begin{subfigure}[b]{0.133\textwidth}
        \centering
        \includegraphics[width=0.98\textwidth]{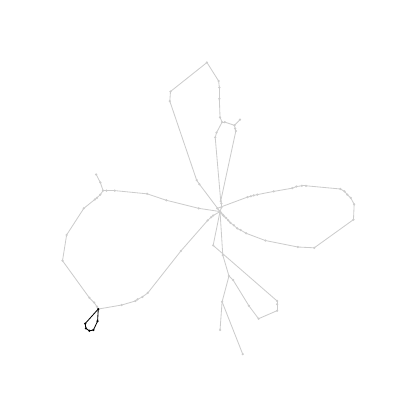}
        \caption{}
        \label{fig:r_ac1}
    \end{subfigure}
    \begin{subfigure}[b]{0.133\textwidth}
        \centering
        \includegraphics[width=0.98\textwidth]{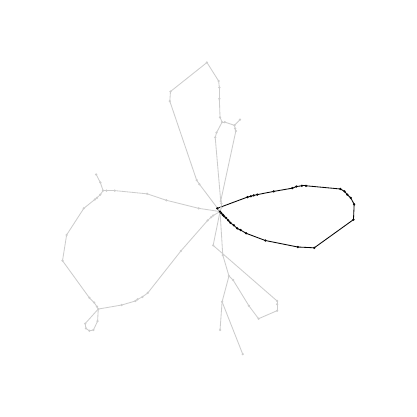}
        \caption{}
        \label{fig:r_ac2}
    \end{subfigure}
    \begin{subfigure}[b]{0.133\textwidth}
        \centering
        \includegraphics[width=0.98\textwidth]{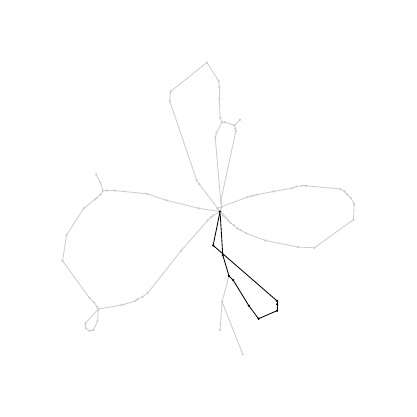}
        \caption{}
        \label{fig:r_ac3}
    \end{subfigure}
    \begin{subfigure}[b]{0.133\textwidth}
        \centering
        \includegraphics[width=0.98\textwidth]{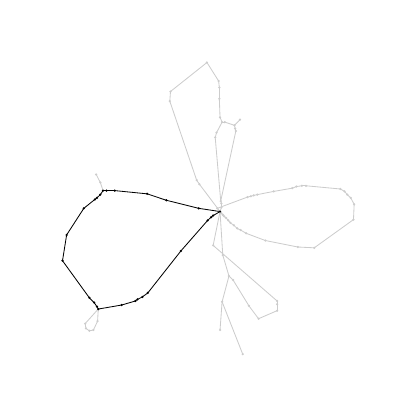}
        \caption{}
        \label{fig:r_ac4}
    \end{subfigure}
    \begin{subfigure}[b]{0.133\textwidth}
        \centering
        \includegraphics[width=0.98\textwidth]{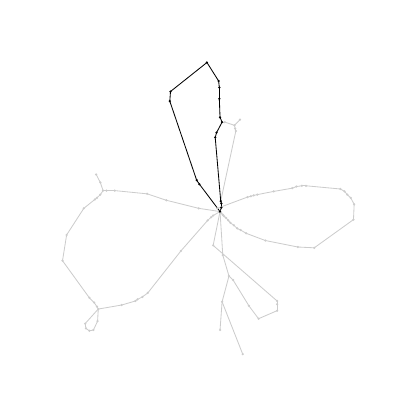}
        \caption{}
        \label{fig:r_ac5}
    \end{subfigure}
    \begin{subfigure}[b]{0.133\textwidth}
        \centering
        \includegraphics[width=0.98\textwidth]{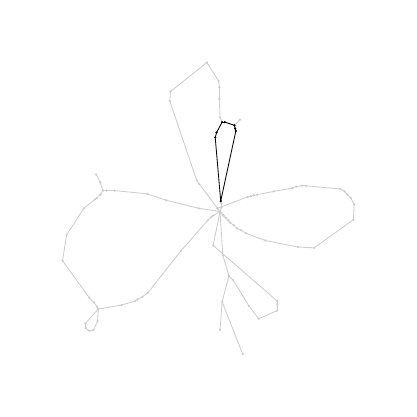}
        \caption{}
        \label{fig:r_ac6}
    \end{subfigure}
    \begin{subfigure}[b]{0.133\textwidth}
        \centering
        \includegraphics[width=0.98\textwidth]{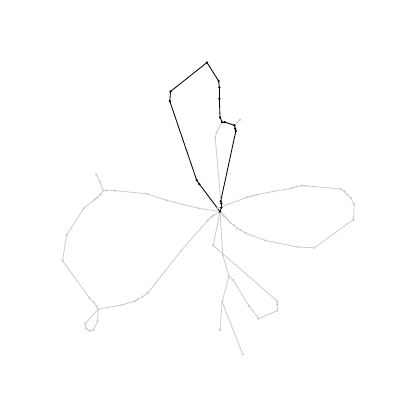}
        \caption{}
        \label{fig:r_ac7}
    \end{subfigure}
\caption{This figure shows all cycles in the \mvr test case enumerated using Gibbs' algorithm \cite{Gibbs69}.}
\label{fig:ruralGibbs}
\end{figure*}
\section{Results}
\label{sec:results}
Computational experiments were conducted on 3 test cases
with 2 different load profiles for two of them, using Juniper \cite{juniper}. \gls{cdsr} and \gls{rrdsr} were solved on \mvr with and without \glspl{res}, creating cases 1 \& 2, as well as on \mvc, yielding cases 3 \& 4 \cite{Meinecke2020}. In \mvc line 38 was switched off additionally in the baseline topology to form a radial network. Furthermore, results for the test case \mvs without \glspl{res} were added, referenced as case 5, which has close to 20 more branches than both other cases and a similar number of acyclicity constraints, compared to \mvc.
\newline
Results are summarized in Table \ref{tab:results}. For each optimization run the baseline topology was used as a MIP start. Without MIP starts, several models could not be solved due to convergence issues. $\Delta p^L=0$ indicates that no improving topology was found, in the time limit.
The superior solver performance on \gls{cdsr} can be explained by tighter search space.
Neglecting \gls{pf}, all spanning subgraphs with $\lvert\V\rvert-1$ edges are considered in \gls{rrdsr} by \eqref{cdsr:nm1}. This must be a super set of all \glspl{st} by Theorem \ref{thm:tree}, which are considered by \eqref{cdsr:nm1} \& \eqref{cdsr:acyclic} in \gls{cdsr}. All variable values are within $9.3\times 10^{-9}$ of the reference \gls{pf} implementation \cite{Coffrin2018} for our \gls{pf} validation instance. For \gls{cdsr}, solved on case 5, a solution is found after approximately 600 s. The solver then times out without additional improvements after 900 s. This shows the need for further research into fast computational methods, especially good heuristics.
\begin{table}[ht]
\renewcommand{\arraystretch}{1.05}
\caption{The results show lower computation times $ct$ for \gls{cdsr}.}
\label{tab:results}
\begin{center}
\resizebox{\columnwidth}{!}{%
\begin{tabular}{c c c c c c c r}
\hline
Case & Model & $\f$ [MW] & $p^L$ [MW] & $\Delta p^L$ [\%] & $\gamma^{v}$ [p.u.] &  $\gamma^{s}$ [MVA] & $ct$ [s]\\
\hline
1 & C-DSR & -8.18 & 0.13 & 33.91 & 0.0 & 0.0 & 19.76\\
1 & RR-DSR &  -8.11  &  0.2  &  0.0  &  0.0  &  0.0  & 900.59\\
2 & C-DSR & 17.51 & 0.25 & 31.76 & 0.0 & 0.0 & 13.52\\
2 & RR-DSR &  17.51 & 0.25 & 31.75 & 0.0104 & 0.0 & 609.47\\
3 & C-DSR & 17.99 & 0.15 & 49.17 & 0.0 & 0.0 & 26.81\\
3 & RR-DSR & 18.14 & 0.29 & 0.0 & 0.0084 & 0.0 & 903.26\\
4 & C-DSR & 34.78 & 0.3 & 38.74 & 0.0045 & 0.0 & 24.76\\
4 & RR-DSR &  34.98  &  0.5  &  0.0  &  0.0259  &  0.0  & 902.35\\
5 & C-DSR & 32.15 & 0.51 & 12.53 & 0.0182 & 0.0 & 906.93\\
5 & RR-DSR &  32.22  &  0.58 &  0.0  &  0.0366  &  0.0  & 902.35\\
\hline
\end{tabular}
}
\end{center}
\end{table}
\setcounter{figure}{1}
\begin{figure}[ht]
    \begin{subfigure}[b]{0.325\linewidth}
        \centering
        \includegraphics[width=0.94\textwidth]{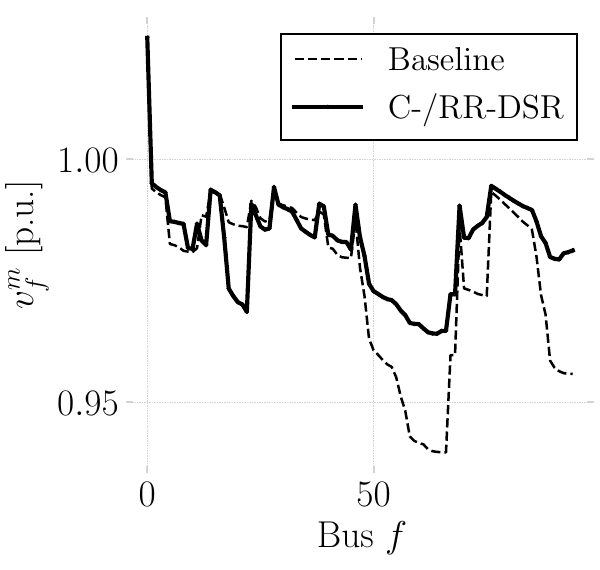}
        \caption{Voltage}
        \label{fig:r_v}
    \end{subfigure}
    \begin{subfigure}[b]{0.325\linewidth}
        \centering
        \includegraphics[width=0.94\textwidth]{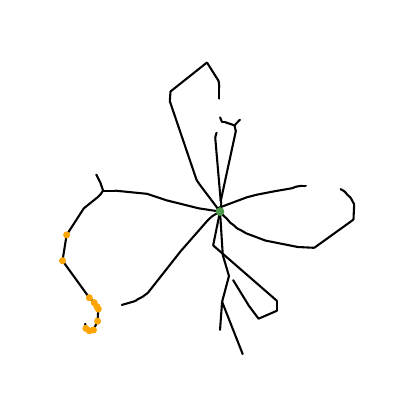}
        \caption{Baseline topology}
        \label{fig:r_b}
    \end{subfigure}
    \begin{subfigure}[b]{0.325\linewidth}
        \centering
        \includegraphics[width=0.94\textwidth]{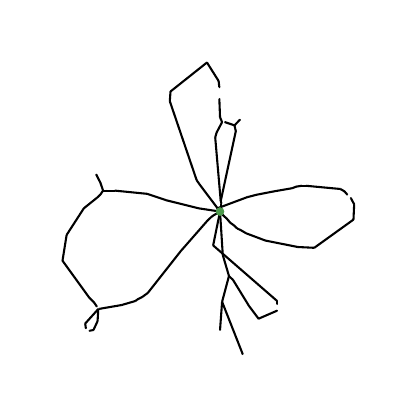}
        \caption{RR-/C-DSR}
        \label{fig:r_opt}
    \end{subfigure}
\caption{The figure shows results for case 2, \mvr without \glspl{res}. In \ref{fig:r_b} \& \ref{fig:r_opt} voltage violations are highlighted in yellow and the substation in green.}
\label{fig:ruralTopos}
\end{figure}
\setcounter{figure}{2}
\begin{figure}[ht]
    \begin{subfigure}[b]{0.325\linewidth}
        \centering
        \includegraphics[width=0.94\textwidth]{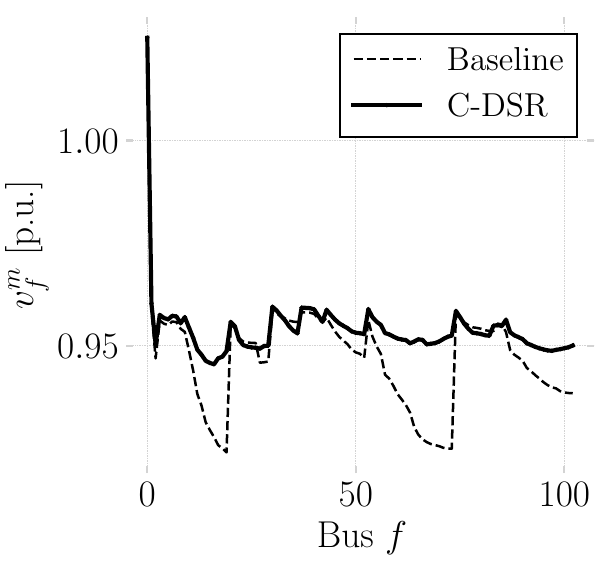}
        \caption{Voltage}
        \label{fig:c_v}
    \end{subfigure}
    \begin{subfigure}[b]{0.325\linewidth}
        \centering
        \includegraphics[width=0.94\textwidth]{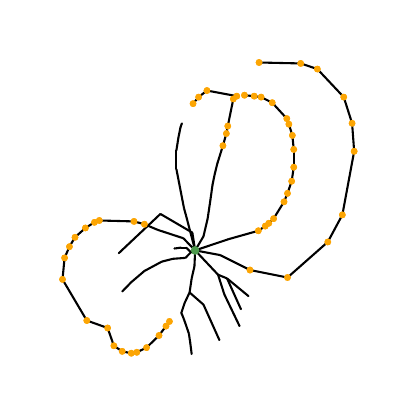}
        \caption{Baseline topology}
        \label{fig:c_b}
    \end{subfigure}
    \begin{subfigure}[b]{0.325\linewidth}
        \centering
        \includegraphics[width=0.94\textwidth]{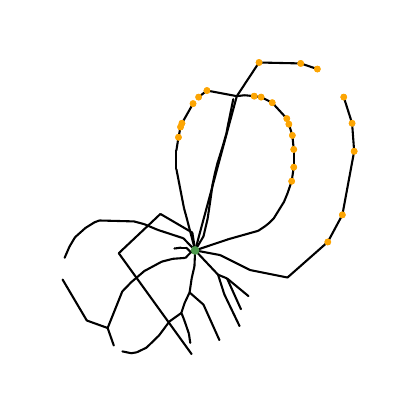}
        \caption{C-DSR}
        \label{fig:c_opt}
    \end{subfigure}
\caption{The figure shows results for case 4, \mvc without \glspl{res}.}
\label{fig:commTopos}
\end{figure}
\setcounter{figure}{2}
\begin{figure}[ht]
    \begin{subfigure}[b]{0.325\linewidth}
        \centering
        \includegraphics[width=0.94\textwidth]{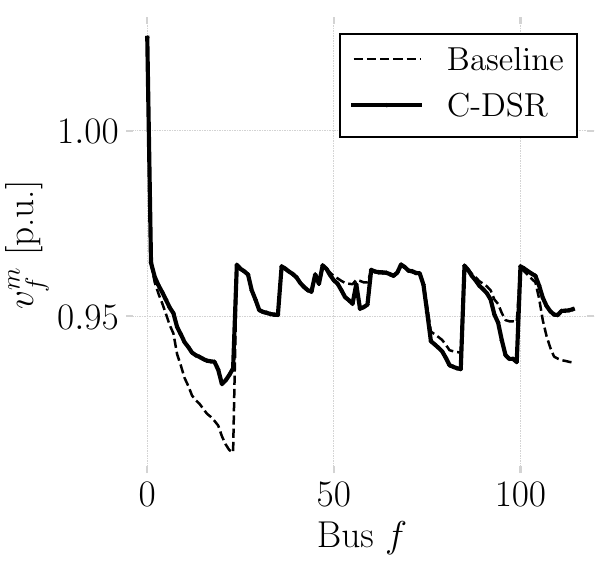}
        \caption{Voltage}
        \label{fig:s_v}
    \end{subfigure}
    \begin{subfigure}[b]{0.325\linewidth}
        \centering
        \includegraphics[width=0.94\textwidth]{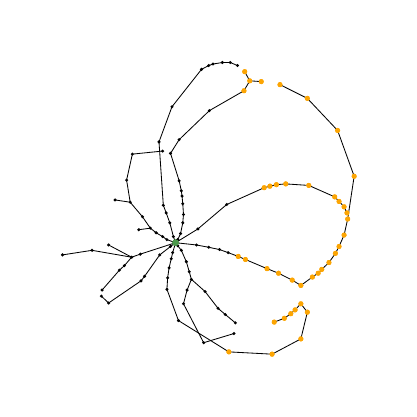}
        \caption{Baseline topology}
        \label{fig:s_b}
    \end{subfigure}
    \begin{subfigure}[b]{0.325\linewidth}
        \centering
        \includegraphics[width=0.94\textwidth]{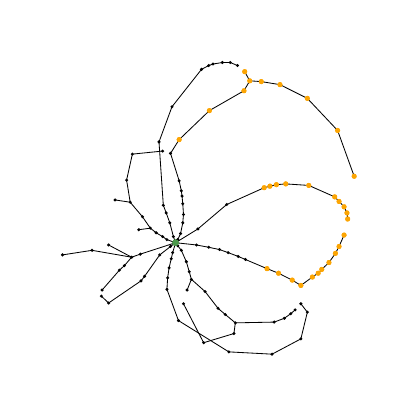}
        \caption{C-DSR}
        \label{fig:s_opt}
    \end{subfigure}
\caption{The figure shows results for case 4, \mvs without \glspl{res}.}
\label{fig:suTopos}
\end{figure}

\section{Conclusion}
\label{sec:conclusion}
The results show that Juniper \cite{juniper} can find solutions for \gls{cdsr} in substantially less time, when compared to \gls{rrdsr}. The additional acyclicity constraints \eqref{cdsr:acyclic} significantly strengthen the formulation. For slightly larger cases, however, the need for fast computational methods already becomes obvious.
The same holds for advanced methods to reduce the number of radiality constraints or additional computational methods that would allow for the optimization of even larger networks.

\bibliographystyle{IEEEtran}
\bibliography{z_bib}

\appendices

\section{Theoretical results}
\label{app:A}
\subsection{Trees, subgraphs and spanning trees}
\label{sec:gt}
All definitions and theorems introduced in Seubsection \ref{sec:gt} are based on \cite{Gross2023}. 
In the following, graph $\N$ is assumed to be connected and undirected.
\begin{theorem}
\label{thm:tree}
Let $\T$ be a graph with $\vts{\V_{\T}}$ vertices. Then the following statements are equivalent.
\begin{enumerate}
\item $\T$ is a tree.
\item $\T$ contains no cycles and has $\vts{\V_{\T}}-1$ edges.
\item $\T$ is connected and has $\vts{\V_{\T}}-1$ edges.
\end{enumerate}
\end{theorem}
\begin{definition}
\label{def:subg}
A subgraph $\H$ is said to span graph $\N$ if $\V_{\N}=\V_{\H}$. $\H$ is then also called a spanning subgraph of $\N$.
\end{definition}
%
\begin{definition}
\label{def:spt}
A spanning tree of a graph is a spanning subgraph that is a tree.
\end{definition}
%
%
%
\begin{definition}
\label{def:cycleEdge}
Edge $e$ of a graph $\N$ is called a \gls{ce} if $e$ lies in a cycle of $\N$.
\end{definition}
\begin{proposition}
\label{prop:conn}
Let $e$ be an edge of a connected graph $\N$. Then $\N-e$ is connected if and only if e is a cycle-edge of $\N$.
\end{proposition}
\subsection{Proof of Corollary \ref{cor:switchable}}
\begin{proof}
If $e\notin\textstyle\bigcup_{\forall k} \EN_k$, e is no \gls{ce} based on Definition \ref{def:cycleEdge}. Removing $e$ from $\N$ disconnects $\N$ by Proposition \ref{prop:conn}.
\end{proof}
\subsection{Proof of Corollary \ref{cor:ces}}
\begin{proof}
Let $\B$ be a \gls{cb} of $\N$. Any cycle in $\N$ can be expressed as the symmetric difference of elements in $\B$ \cite{Deo2016}. Assume $\exists e\in\E^{N}_i$, for which $e\notin\E(\CF_k),~\forall k\in\{1,\dots,\beta(\N)\}$. Then, $\B$ is clearly no basis of the cycle space of $\N$, which is a contradiction.
\end{proof}

\section{Set of enumerated cycles of \mvr}
\label{app:B}


\end{document}